\documentclass[twocolumn,prb,aps,amsmath,superscriptaddress]{revtex4}
\usepackage{graphicx}
\usepackage{xspace} 
\usepackage{multirow} 
\usepackage{dcolumn} 

\newcommand{\jc}{\ensuremath{J_\mathrm{c}}\xspace}
\newcommand{\ic}{\ensuremath{I_\mathrm{c}}\xspace}
\newcommand{\jb}{\ensuremath{J_\mathrm{b}}\xspace}
\newcommand{\jp}{\ensuremath{J_{\mathrm{b}\bot}}\xspace}

\newcommand{\Ic}{\ic{}}

\newcommand{\dg}{\ensuremath{^\circ}\xspace}
\newcommand{\degrees}{\dg{}}

\renewcommand{\vec}{\mathbf}

\begin{document}

\title{Enhanced current flow through meandering and tilted grain boundaries in YBCO films}
\author{Rafael B. Dinner}
\author{Kathryn A. Moler}
\email[Electronic address: ]{kmoler@stanford.edu}
\affiliation{Geballe Laboratory for Advanced Materials, Stanford University, Stanford, California 94305}
\author{D. Matthew Feldmann}
\affiliation{Los Alamos National Laboratory, Los Alamos, New Mexico 87545}
\author{M. R. Beasley}
\affiliation{Geballe Laboratory for Advanced Materials, Stanford University, Stanford, California 94305}

\begin{abstract}
Grain boundaries (GBs) have been shown to limit critical current density, \jc, in YBa$_2$Cu$_3$O$_{7-\delta}$ (YBCO) coated conductors. Here we use transport measurements and scanning Hall probe microscopy coupled with current reconstruction to demonstrate that GB geometry, such as the in-plane meandering observed in films grown by metalorganic deposition (MOD) on rolling assisted biaxially textured substrate (RABiTS), can lead to higher GB \jc. We observe current-induced flux entry into such a coated conductor, then model its behavior by imaging films with single, straight GBs tilted at various angles to the applied current.
\end{abstract}

\maketitle

High-temperature superconductors are being engineered for use in high-current applications, including motors, generators, and power lines.\cite{bes,400Hz} Present efforts center around the coated conductor architecture, in which a YBa$_2$Cu$_3$O$_{7-\delta}$ (YBCO) film is grown on a flexible metal tape. The tape is polycrystalline, and its crystal grains give rise to separate grains in the superconducting layer. At the grain boundaries (GBs), superconductivity is weakened, and magnetic flux, in the form of vortices, penetrates at lower critical current densities than those supported by the intragranular YBCO.\cite{dimos_gb,hts_gb_rmp,feldmann_gb}

\begin{figure}
 \begin{center}
  \includegraphics[width=6.5cm]{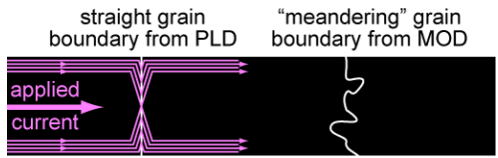}
  \caption{\label{meandering_fig}While PLD-grown YBCO replicates the straight grain boundaries of the substrate, MOD leads to meandering boundaries, which have been found to support higher critical currents.\protect\cite{feldmann_meandering} The current streamlines show a critical state prediction for current spreading along a straight boundary.}
 \end{center}
\end{figure}

In previous work,\cite{feldmann_meandering} films grown on rolling assisted biaxially textured substrates (RABiTS) by two different methods, metalorganic deposition (MOD) and pulsed laser deposition (PLD), were found to have a similar \textit{intragrain} critical current density \jc (4.5 versus 5.1 MA/cm$^2$ at 0~T, 77~K), but rather different \textit{intergrain} critical current density \jb (0.9 versus 3.4 MA/cm$^2$), as computed using the sample cross sectional area. Grain mapping of the MOD film's substrate compared to layers in the YBCO found two possible contributions to its enhanced \jb. First, the grain alignment improved from the substrate to the YBCO. Second, the YBCO GBs meander relative to the straight substrate GBs, as depicted in Fig.~\ref{meandering_fig}. In contrast, PLD films replicate the substrate grains' orientations and straight GBs.

\begin{figure}
 \begin{center}
  \includegraphics[width=8.5cm]{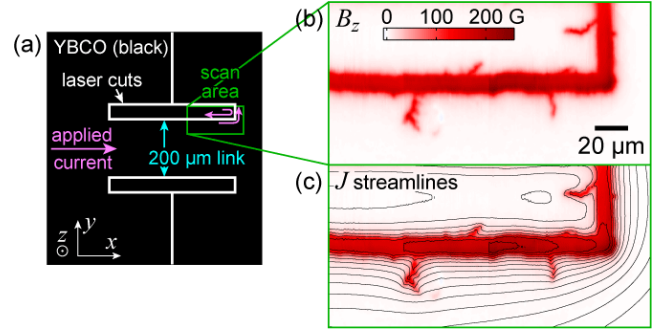}
  \caption{\label{amsc}(a) A sketch of an American Superconductor YBCO coated conductor grown by MOD on RABiTS, in which laser cuts through the film define a bridge that channels the applied current. (b) Magnetic field, $B_z$, over a section of the bridge, imaged via scanning Hall probe microscopy. A 1.3~A applied current forces flux into the film from the cut edge, primarily along narrow, meandering features presumed to be grain boundaries. (c) Reconstructed current streamlines overlaid on $B_z$ show that the current spreads out along the boundaries, and its direction varies to accommodate their shape. 40~mA flows between neighboring streamlines. The full set of images is presented as a movie in the supplemental material.\protect\cite{epaps}}
 \end{center}
\end{figure}

Here, we use magnetic imaging and transport measurements to confirm that meandering grain boundaries can enhance \jb. Meandering increases the length (and hence area) of the GB connecting two grains of a given width. If current redirects to spread along the GB, it can take advantage of this length and \ic may thereby increase proportionally.

The magnetic images in this work are obtained with a custom-built large-area scanning Hall probe microscope.\cite{rsi} The sample is held at 40~K in vacuum and a sinusoidal current is applied. A Hall sensor rasters over the sample surface, pausing at each pixel location to measure $B_z(t)$ over the full cycle of applied current. The data is later assembled into a series of images, each corresponding to a particular phase---and applied current---within the cycle.

\begin{figure*}
 \begin{center}
  \includegraphics[width=\textwidth]{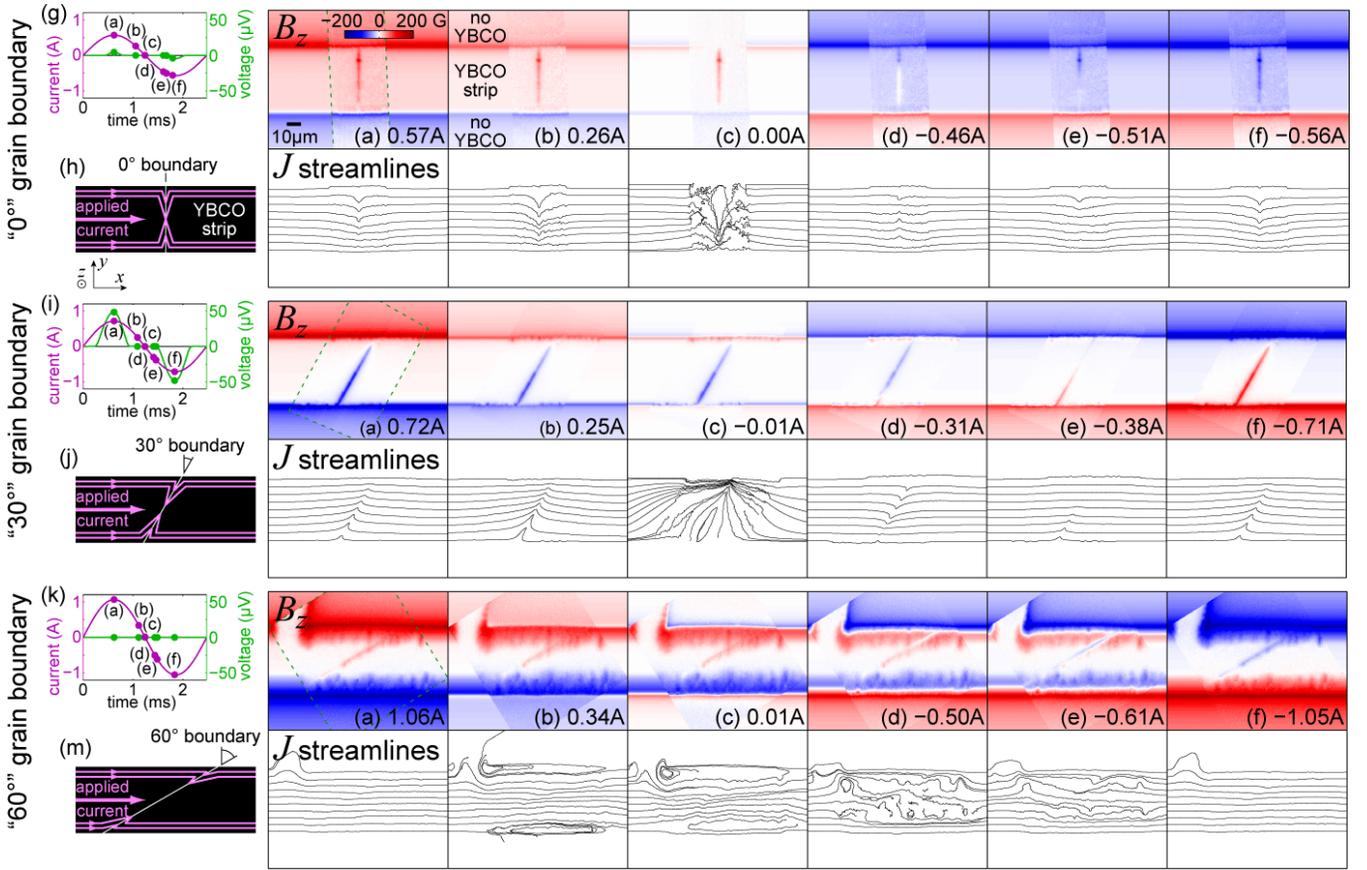}
  \caption{\label{3060}(a)--(f) Magnetic images, $B_z$, and reconstructed streamlines of current, $J$, for each of three samples with grain boundaries tilted as shown in (h), (j), and (m). The boundary between data and background calculation is identified in the (a) frames by a dashed green box. The magnetic images are all on the same color scale, shown in (a). The full set of images is presented as a movie in the supplemental material.\protect\cite{epaps} (g), (i), and (k) plot the applied currents and measured voltages for frames (a)--(f) within the current cycle.}
 \end{center}
\end{figure*}

We start by imaging an MOD film---a 400~nm thick YBCO coated conductor grown by American Superconductor on RABiTS. Figure~\ref{amsc}a shows the sample geometry, defined by lines cut through the film with a Nd-YAG laser. The cuts direct the applied current to flow through a 200~$\mu$m wide by 450 ~$\mu$m long link. The link is held at 40~K and carries 1.3~A of current at the peak of an ac cycle. As depicted by the pink arrows within the scan area in Fig.~\ref{amsc}(a), the current flows along the link at the bottom edge of the laser cut, and shielding current flows around the isolated section of YBCO that constitutes its upper edge, in response to the field generated by the link. These currents force flux into the film from the edges, primarily along the narrow, crack-like features seen in the magnetic image, Fig.~\ref{amsc}(b). Combined magneto-optic and grain mapping studies of similar samples have shown that such narrow features are GBs.\cite{feldmann_gb} Our image reveals that the GBs are not straight. By reconstructing the current density, $\vec{J}$, from the magnetic image,\cite{E,reg} we show in Fig.~\ref{amsc}(c) that $\vec{J}$ does change direction to accommodate the shape of the meandering GBs, at least down to the length scales of the instrument resolution of about 1~$\mu$m.

To model the behavior of a local segment of a meandering GB, we fabricate straight GBs placed at various in-plane angles across links of fixed width (50~$\mu$m), depicted in Fig.~\ref{3060}(h), (j), and (m). The GB length increases as it tilts toward the direction of current flow between the grains. We pattern the samples from two nominally identical 180~nm thick YBCO films, grown epitaxially by pulsed laser deposition on 5\degrees{} miscut SrTiO$_3$ [001] bicrystal substrates, each with a straight GB. The 0\dg link, Fig.~\ref{3060}(h), occupies one film, while the 30\dg and 60\dg links are patterned at different points along the GB of the second film.

\begin{table}
		\begin{tabular}{r|dd|ddd}
		   \multirow{2}{*}{sample} &
		   \multicolumn{2}{c|}{\jb (MA/cm$^2$)} &
		   \multicolumn{3}{c}{\jp (MA/cm$^2$)} \\
		   
		      &
		   \multicolumn{1}{c}{\hspace{.5em} 77 K \hspace{.5em} } &
		   \multicolumn{1}{c|}{40 K} &
		   \multicolumn{1}{c}{\hspace{.3em} 77 K \hspace{.3em} } &
		   \multicolumn{1}{c}{40 K} &
		   \multicolumn{1}{c}{\hspace{.2em} CR \hspace{.2em}} \\
		   \hline
		   
			 0\dg GB & 1.4 &     5.6 & 1.4 &    5.6 & 4.6 \\
			30\dg GB & 1.8 &     4.9 & 1.6 &    4.3 & 3.5 \\
			60\dg GB & 4.7 &   >11.8 & 2.4 &   >5.9 & 4.5 \\
			no GB    & 4.7 &    12.6 &     &        &     \\
		\end{tabular}
	\caption{\label{transport_table}Zero-field transport measurements of critical current across three grain boundaries (GBs) normalized by the cross sectional area of the strip (\jb) or the GB (\jp). The last row compares the intragrain \jc. The ``CR'' column lists current density perpendicular to the GBs as reconstructed from magnetic images at 40~K.}
\end{table}

We first measure current-voltage characteristics of the links at 77~K in liquid nitrogen with a critical voltage criterion of 50~nV. The zero-field results are shown in Table~\ref{transport_table}, where we calculate grain boundary critical current density using the strip cross sectional area for \jb, and the GB area for \jp (the current density flowing perpendicular to the GB). Instead of a constant \jp, we find it increases with GB tilt, exceeding the expected geometric effect of lengthening the GB. In fact, at 60\dg tilt, \jb is equal to the measured intragrain value, as if the GB were completely transparent to the current. This strong dependence of \jb on tilt motivates our more detailed study of these samples via magnetic imaging.

Noise characteristics of the Hall sensor force us to image at 40~K rather than 77~K. We therefore also measure current-voltage characteristics in the scanning Hall probe cryostat (at 40~K in vacuum) and include the results in Table~\ref{transport_table}. We were not able to exceed the 50~nV criterion for the 60\degrees{} GB, so we can only place a lower bound on \jb. We also note that at 40~K, the 30\dg \jb values, compared to the 0\dg and 60\dg, are unexpectedly low. These values held at lower amplitudes of ac applied current, so they likely do not arise from sample heating.

Scanning Hall probe images, presented in Fig.~\ref{3060}, yield a more complete picture of how the grain boundaries break down with current. Each of the image series tracks changes in $B_z$ as the sinusoidal applied current decreases from maximum to minimum. The area surrounding each image is filled in with a critical state calculation.\cite{E} From each $B_z$ image, we reconstruct a map of $\vec{J}$ in the film.\cite{E,reg}

The GBs' limiting effect on critical current is immediately evident in the magnetic images, where vortices fully penetrate the GBs but not the grains, even for the 60\dg sample where transport saw no effect of the GB. With increasing GB tilt, however, we must apply successively greater currents to force flux across the GBs, leading to greater flux penetration into the grains.

The $J$ streamlines reveal an abrupt change in direction at each boundary. This is a consequence of the current spreading to lower its density when crossing the GB, and is expected from a critical state model,\cite{mo_gb} whose current streamlines are shown in Fig.~\ref{3060}(h), (j), and (m). One contrast with the model is that the observed flux enters asymmetrically: predominantly from the top of the 0\degrees{} boundary, the bottom of the 30\degrees{}, and the top of the 60\degrees{}. These asymmetries are reproducible over many cycles of applied current, thus we believe that they stem from real, though not necessarily large, asymmetries in the samples.

From the reconstructed $\vec{J}$, we can directly extract \jp, the component of $\vec{J}$ perpendicular to the grain boundary (Table~\ref{transport_table} column 6), for comparison with transport. We average over pixels in a region around the GB that we expect to have reached the intragrain \jc, from the critical state model. \jp is consistent between the 0\dg and 60\dg samples, which contrasts with the transport results, supporting the picture of \jb proportional to GB length and showing that the 60\dg critical current is still below that of the intragrain material. The 30\dg \jp, as in the 40~K transport results, is anomalously low.

The magnetic images offer some insight into why transport yields higher \jb values, particularly for the 60\dg GB. First, we estimate the voltage accompanying \jp from the product of the rate at which the flux front moves along the GB and the density of vortices along the GB. This yields $\sim$10~nV, which is less than the transport \jb criterion of 50~nV, so we expect current reconstruction to yield lower \jp than transport. Second, the density of vortices along the 60\dg GB is lowest---6 vortices/$\mu$m compared to 11 and 10 for the 0\dg and 30\dg GBs. Thus to achieve a fixed voltage criterion, the vortices must flow faster along the 60\dg GB, requiring a higher \jp.

In summary, both transport and magnetic measurements demonstrate that grain boundaries admit higher currents as they tilt toward the direction of current flow. This geometric effect contributes to the transparency of meandering grain boundaries, found in MOD coated conductors on RABiTS. Though transport indicates that \Ic{} increases faster than the grain boundary length, magnetic imaging shows that this is not the case at lower voltage criteria.

%

We gratefully acknowledge American Superconductor Corp. and George Daniels for YBCO, Janice Guikema for Hall sensors, David Kisker for 2DEG, and Joseph Sulpizio and Hung-Tao Chou for oxide growth. This work is funded by the Air Force Multi-University Research Initiative (MURI).


\begin{thebibliography}{10}

\bibitem{bes}
Report of the US DOE Office of Basic Energy Sciences,
\newblock {\em Basic Research Needs for Superconductivity}, 2006.

\bibitem{400Hz}
S.~Kalsi,
\newblock Proc. IEEE {\bf 92}, 1688 (2004).

\bibitem{dimos_gb}
D.~Dimos, P.~Chaudhari, and J.~Mannhart,
\newblock Phys. Rev. B {\bf 41}, 4038 (1990).

\bibitem{hts_gb_rmp}
H.~Hilgenkamp and J.~Mannhart,
\newblock Rev. Mod. Phys. {\bf 74}, 485 (2002).

\bibitem{feldmann_gb}
D.~M. Feldmann et~al.,
\newblock Appl. Phys. Lett. {\bf 77}, 2906  (2000).

\bibitem{feldmann_meandering}
D.~M. Feldmann et~al.,
\newblock J. Mater. Res. {\bf 21}, 923  (2006).

\bibitem{epaps}
Movies and other supplementary materials are available at \url{http://www.stanford.edu/group/moler/rdinner}

\bibitem{rsi}
R.~B. Dinner, M.~R. Beasley, and K.~A. Moler,
\newblock Rev. Sci. Instrum. {\bf 76}, 103702  (2005).

\bibitem{E}
R.~B. Dinner, D.~M. Feldmann, K.~A. Moler, and M.~R. Beasley,
\newblock Imaging ac losses in superconducting films via scanning {Hall} probe
  microscopy,
\newblock Submitted to \textit{Phys. Rev. B}., 2006.

\bibitem{reg}
D.~M. Feldmann,
\newblock Phys. Rev. B {\bf 69}, 144515 (2004).

\bibitem{mo_gb}
A.~A. Polyanskii et~al.,
\newblock Phys. Rev. B {\bf 53}, 8687  (1996).

\end{thebibliography}
\end{document}